\documentclass{LMCS}

\def\doi{9(2:08)2013}
\lmcsheading%
{\doi}
{1--22}
{}
{}
{Aug.~31, 2012}
{Jun.~21, 2013}
{}

\usepackage[utf8x]{inputenc}
\usepackage{MnSymbol}
\usepackage{tikz}                         
\usetikzlibrary{automata}
\usetikzlibrary{positioning}
\usetikzlibrary{arrows}
\usepackage{url}
\usepackage{booktabs}
\usepackage{multirow}
\usepackage{amsmath, amsthm,hyperref,enumerate}

\renewcommand{\phi}{\varphi}

\newcommand{\Aut}[1]{\mathbf{#1}}
\newcommand{\Quot}[2]{#1/{#2}}
\newcommand{\Inf}[1]{\mathop{\mathrm{inf}}(#1)}
\newcommand{\InfK}[2]{\mathop{\mathrm{inf}_{#2}}(#1)}
\newcommand{\Occ}[1]{\mathop{\mathrm{occ}_1}(#1)}
\newcommand{\Occk}[1]{\mathop{\mathrm{occ}_k}(#1)}
\newcommand{\OccK}[2]{\mathop{\mathrm{occ}_{#2}}(#1)}
\newcommand{\Prfk}[1]{\mathop{\mathrm{prf}_k}(#1)}

\newcommand{\Infk}[1]{\mathop{\mathrm{inf}_k}(#1)}
\newcommand{\Sffxk}[1]{\mathop{\mathrm{sffx}_k}(#1)}

\newcommand{\Loop}[1]{{#1{\rcurvearrowup}}}

\newcommand{\No}[2]{|#2|_{#1}}

\newcommand{\AutA}{{\boldsymbol A}}
\newcommand{\AutPhi}{\AutA_{\phi}}
\newcommand{\TL}[1]{\text{TL}_{#1}}

\newcommand{\TLFrag}[2]{\TL{#1} [#2]}

\newcommand{\QuotA}{\Quot{\AutA}{\LeftCon}}
\newcommand{\ClosA}{\mathrm T(\AutA)}
\newcommand{\ClosQA}{\mathrm T(\Quot \AutA {\equiv_\AutA})}
\newcommand{\ClosPhi}{\mathrm T(\Quot \AutPhi {\equiv_{\AutPhi}})}

\newcommand{\Next}{\mathsf{X}}
\newcommand{\Eventually}{\mathsf{F}}
\newcommand{\StrictEvent}{\mathsf{X}\hspace*{-0.2em}\mathsf{F}}
\newcommand{\StrictAlw}{\mathsf{X}\hspace*{-0.3em}\mathsf{G}}
\newcommand{\Always}{\mathsf{G}}
\newcommand{\Until}[2]{#1 \mathsf{U} #2}
\newcommand{\UntilSym}{\mathsf{U}}
\newcommand{\Release}[2]{#1 \mathsf{R} #2}
\newcommand{\ReleaseSym}{\mathsf{R}}
\newcommand{\Naturals}{\mathbf{N}}
\newcommand{\Sub}[1]{\text{sub}(#1)}

\newcommand{\Lan}[1]{\mathrm{L}(#1)}
\newcommand{\LL}[1]{\mathrm{LL}(#1)}

\newcommand{\LeftCon}{\equiv_\AutA}
\newcommand{\NotLeftCon}{\not\equiv_\AutA}
\newcommand{\NZG}{\nleftrightarrow}

\tikzstyle{every node}=[circle, inner sep=1pt, outer sep=2pt, minimum width=5mm, node distance=15mm, semithick, initial text={}]
\tikzstyle{lab}=[minimum width=0mm,outer sep=2pt,rectangle]
\tikzstyle{every picture}=[|-]

\begin{document}

\begin{abstract}
  We present a framework for obtaining effective characterizations of simple fragments of future temporal logic (LTL) with the natural numbers as time domain. The framework is based on a form of strongly unambiguous automata, also known as prophetic automata or complete unambiguous Büchi automata and referred to as Carton--Michel automata in this paper. These automata enjoy strong structural properties, in particular, they separate the ``finitary fraction" of a regular language of infinite words from its ``infinitary fraction" in a natural fashion. Within our framework, we provide characterizations of several natural fragments of temporal logic, where, in some cases, no effective characterization had been known previously, and give lower and upper bounds for their computational complexity.
\end{abstract}

\title[Effective Characterizations of Simple Fragments of Temporal Logic]{Effective Characterizations of Simple Fragments of Temporal Logic Using Carton--Michel Automata}

\author[S.~Preugschat]{Sebastian Preugschat}
\address{Christian-Albrechts-Universität zu Kiel, 24098 Kiel, Germany}
\email{\{preugschat,wilke\}@ti.informatik.uni-kiel.de}

\author[Th.~Wilke]{Thomas Wilke}
\address{\vskip-6 pt}

 \ACMCCS{[{\bf Theory of computation}]: Formal
  languages and automata theory---Automata over infinite objects;
  Logic---Modal and temporal logics }

\maketitle

\section{Introduction}

Ever since propositional linear-time temporal logic (LTL) was introduced into computer science by Amir Pnueli in~\cite{pnueli-focs-1977} it has been a major object of research. The particular line of research we are following here is motivated by the question how each individual temporal operator contributes to the expressive power of LTL. More precisely, our objective is to devise decision procedures that determine whether a given LTL property can be expressed using a given subset of the set of all temporal operators, for instance, the subset that includes ``next'' and ``eventually'', but not ``until''.

As every LTL formula interpreted in the natural numbers (the common time domain) defines a regular language of infinite words ($\omega$-language), the aforementioned question can be viewed as part of a larger program: classifying regular $\omega$-languages, that is, finding effective characterizations of subclasses of the class of all regular $\omega$-languages. Over the years, many results have been established and specific tools have been developed in this program, the most fundamental result being the one that says that a regular $\omega$-language is star-free or, equivalently, expressible in first-order logic or in LTL if, and only if, its syntactic semigroup is aperiodic~\cite{kamp-1968,thomas-ic-1979,perrin-mfcs-1984}.

The previous result is a perfect analogue of the same result for regular languages of finite words, that is, of the classical theorems by Schützenberger~\cite{schuetzenberger-ic-1965}, McNaughton and Papert~\cite{mcnaughton-papert-1971}, and Kamp~\cite{kamp-1968}. In general, the situation with infinite words is more complicated than with finite words; a good example for this is given in~\cite{diekert-kufleitner-mst-2011}, where, for instance, tools from topology and algebra are used to settle characterization problems for $\omega$-languages.

The first characterization of a fragment of LTL over finite linear orderings was given in \cite{cohen-perrin-pin-jcss-1993}, another one followed in \cite{etessami-wilke-ic-2000}, both following a simple and straightforward approach: to determine whether a formula is equivalent to a formula in a certain fragment, one computes the minimum reverse DFA for the corresponding regular language and verifies certain structural properties of this automaton, more precisely, one checks whether certain ``forbidden patterns'' do not occur. The first characterization for infinite words (concerning stutter-invariant temporal properties) \cite{peled-wilke-wolper-tcs-1998} used sequential relations on $\omega$-words; the second (concerning the nesting depth in the until/since operator) \cite{wilke-habil-1998} used heavy algebraic machinery and did not shed any light on the computational complexity of the decision procedures involved. In fact, the upper bound that can be derived from this work is non-elementary.

In this paper, we describe a general, conceptually simple paradigm for characterizing fragments of LTL when interpreted in the natural numbers, combining ideas from \cite{cohen-perrin-pin-jcss-1993,etessami-wilke-ic-2000} for finite words with the work by Carton and Michel on unambiguous Büchi automata \cite{carton-michel-latin-2000,carton-michel-tcs-2003}. The approach works roughly as follows. To determine whether a given formula is equivalent to a formula in a given fragment, convert the formula into what is called a ``prophetic automaton'' in \cite{perrin-pin-elsevier-2004}, check that the automaton, when viewed as an automaton on finite words, satisfies certain properties, and check that languages of finite words derived from the accepting loops (``loop languages'') satisfy certain other properties. In other words, we reduce the original problem for $\omega$-languages to problems for languages of finite words. We show that the approach works for all reasonable fragments of future LTL and yields optimal 
upper bounds for the complexity of the corresponding decision procedures for all but one fragment.

Clearly, the prophetic automaton we start out with is the output of a straightforward translation; one cannot (!) expect that it provides much information about the nature of the language recognized. When we check properties of the automaton when viewed as an automaton on finite words, we first take a quotient, which makes the automaton in some sense canonical. In addition, when we derive the loop languages (representing the infinitary part of the given language) we do this with respect to that quotient, making the loop languages canonical in some sense. This approach ensures that overall we do not analyze more or less arbitrary objects derived from the given formula, but objects (languages) representing very well the nature of the property defined.

Fragments of temporal logic have been studied from different perspectives. One question that has been raised several times is what exactly is the right fragment to specify a given system. A very general answer to this has been given by Leslie Lamport in his seminal paper, \cite{lamport-1994}, on the Temporal Logic of Actions. Another question that has been worked on is how the complexity of model checking depends on the particular fragment considered; results on this can already be found in the groundbreaking paper \cite{sistla-clarke-jacm-1985}, by A.\ Prasad Sistla and Edmund Clarke. The perspective taken in this paper is different, as pointed out above.

\paragraph*{A note on terminology.} 

As just explained, we work with a variant (for details, see below) of the automaton model introduced by Carton and Michel in \cite{carton-michel-latin-2000,carton-michel-tcs-2003} and named CUBA model (\textbf{C}omplete \textbf{U}nambiguous \textbf{B}üchi \textbf{A}utomata). In \cite{perrin-pin-elsevier-2004}, Pin uses ``prophetic automata'' to refer to CUBA's. In the conference version of this paper, \cite{preugschat-wilke-2012}, we referred to these automata as ``Carton--Michel automata'' (CMA) and we stick to this terminology in this paper. At the conference, STACS~2012, Thomas Colcombet gave an invited talk on determinism, non-determinism, and unambiguity with a very broad perspective and used, justified by his broad perspective, the notion ``strongly unambiguous automata'' (SUA) for a somewhat weaker form of unambiguity, see also the contribution to the conference proceedings, \cite{colcombet-2012}. 

\paragraph*{Outline.} 

In Section~2, we provide background on the topics relevant to this paper, in particular, CMA's, propositional linear-time temporal logic, and its translation into CMA's. In Section~3, we present our characterizations. In Section~4 to Section~8, we give proofs of the correctness of our characterizations, and in Section~9, we explain how our characterizations can be used effectively and deal with complexity issues. We conclude with open problems.


\section{Basic Notation and Background}

\subsection{Reverse Deterministic Büchi Automata}

A \emph{Büchi automaton with a reverse deterministic transition function} is a tuple $(A, Q, I, \cdot, F)$ where
\begin{enumerate}[$-$]
\item $A$ is a finite set of symbols,
\item $Q$ is a finite set of states, 
\item $I \subseteq Q$ is a set of initial states, 
\item $\cdot$ is a reverse transition function $A \times Q \to Q$, and
\item $F \subseteq Q$ is a set of final states.
\end{enumerate}
As usual, the transition function is extended to finite words by setting $\epsilon \cdot q = q$ and $au \cdot q = a \cdot (u \cdot q)$ for $q \in Q$, $a \in A$, and $u \in A^*$. For ease in notation, we write $uq$ for $u \cdot q$ when the transition function $\cdot$ is clear from the context. 

A run of an automaton as above on an $\omega$-word $u$ over $A$ is an $\omega$-word $r$ over~$Q$ satisfying the condition $r(i) = u(i) r(i+1)$ for every $i < \omega$. Such a run is called \emph{initial} if $r(0) \in I$; it is \emph{final} if there exist infinitely many $i$ such that $r(i) \in F$; it is \emph{accepting} if it is initial and final. The language of $\omega$-words recognized by such an automaton, denoted $\Lan{\AutA}$ when $\AutA$ stands for the automaton, is the set of $\omega$-words for which there exists an accepting run.

\subsection{Carton--Michel Automata}

An automaton as above is called a \emph{Carton--Michel automaton (CMA)} if for every $\omega$-word over $A$ there is exactly one final run. Such an automaton is \emph{trim}, if every state occurs in some final run.--- The original definition of Carton and Michel in \cite{carton-michel-latin-2000,carton-michel-tcs-2003} is slightly different, but for trim automata---the interesting ones---the definitions coincide.

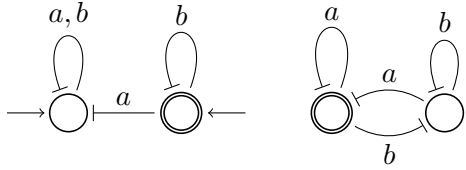
\begin{figure}[t]%
\centering
\begin{tikzpicture}
  \node[initial, draw] (0)                                {};
  \node[initial, initial where=right, accepting, outer sep=3pt, draw] (1) [right of=0]                   {};
  \node[accepting, outer sep=3pt, draw] (2) [node distance=20mm,right of=1]                   {};
  \node[draw] (3) [right of=2]                   {};
  \path 	  
  (0) edge node[lab,above]{$a$} (1);
  \path
  (0) edge [loop, looseness=12,out=110,in=70] node[lab,above]{$a,b$} (0);
  \path
  (1) edge [loop, looseness=11,out=110,in=70] node[lab,above]{$b$} (1);
  \path
  (2) edge [loop, looseness=12,out=110,in=70] node[lab,above]{$a$}
  (2) edge [bend left] node[lab,above]{$a$} (3)
  (3) edge [bend left] node[lab,below]{$b$} (2)
  (3) edge [loop, looseness=11, out=110, in=70] node[lab,above]{$b$}(3) 
  ;
\end{tikzpicture}
\caption{CMA which recognizes $(a+b)^*b^\omega$}%
  \label{fig:cuba}
\end{figure}

As an example, consider the automaton depicted in Figure~\ref{fig:cuba}, which is a CMA for the language denoted by $(a+b)^*b^\omega$. Note that we depict $p = aq$ as
\begin{center}
\begin{tikzpicture}
  \node[draw] (0) {$p$};
  \node[draw,right of=0] (1) {$q$};
  \path[|-] (0) edge node[lab,above]{$a$} (1); 
\end{tikzpicture} 
\end{center}

An initial state has an incoming edge \raisebox{0.3em}{\tikz \draw[->] (0,0) -- (0.5,0);}, a final state has a double circle \tikz \node[accepting,draw,minimum width=3mm]{};. Note that both components in Figure~\ref{fig:cuba} belong to the automaton. The right component is needed to satisfy the condition that every $\omega$-word has a final run in the automaton.

The fundamental result obtained by Carton and Michel is the following.

\begin{thm}[Carton and Michel \cite{carton-michel-latin-2000,carton-michel-tcs-2003}]
  \label{thm:cm}
  Every regular $\omega$-language is recognized by some CMA. More precisely, every Büchi automaton with $n$ states can be transformed into an equivalent CMA with at most $(12n)^n$ states.
\end{thm}

Let $\AutA$ be a CMA over an alphabet $A$ and $u \in A^+$. The word $u$ is a \emph{loop at $q$} if $q = uq$ and there exist $v, w \in A^*$ satisfying $vw = u$ and $w q \in F$. The set of loops at $q$ is denoted $S(q)$. What Carton and Michel prove about loops is: 

\begin{lem}[Carton and Michel \cite{carton-michel-latin-2000,carton-michel-tcs-2003}]
  \label{lem:anchor}
  Let $\AutA$ be a CMA over some alphabet~$A$. Then, for every $u \in A^+$, there is exactly one state $q$, denoted $\Loop u$ and called \emph{anchor of $u$}, such that $u$ is a loop at $q$.

  In other words, the $S(q)$'s are pairwise disjoint and $\bigcup_{q \in Q}S(q) = A^+$.
\end{lem}

\subsection{Generalized Carton--Michel Automata}

A \emph{generalized Carton--Michel automaton (GCMA)} is defined as expected. It is the same as a CMA except that the set $F$ of final states is replaced by a set $\mathfrak F \subseteq 2^Q$ of final sets, just as with ordinary generalized Büchi automata. For such an automaton, a run $r$ is final if for every $F \in \mathfrak F$ there exist infinitely many $i$ such that $r(i) \in F$.

The above definitions for CMA's can all be adapted to GCMA's in a natural fashion. For instance, a word $u$ is a loop at some state $q$ in a GCMA if $q = uq$ and for every $F \in \mathcal F$ there exist $v,w \in A^*$ such that $u = vw$ and $wq \in F$. 

It is a theorem by Carton and Michel that every GCMA can be converted into an equivalent CMA:

\begin{thm}[Carton and Michel \cite{carton-michel-latin-2000,carton-michel-tcs-2003}]
  \label{thm:cm2}
  Let $\Aut A = (A, Q, I, \cdot, \mathfrak F)$ be a GCMA such that $|Q|=n$ and $|\mathfrak F| = m$. There is an equivalent CMA $\Aut A' = (A', Q', I', \cdot', \mathfrak F')$ such that $|Q'| \leq 2^{mn}$.
\end{thm}

The proof of Lemma~\ref{lem:anchor} given in \cite{carton-michel-tcs-2003} carries over to GCMA's without any change. Therefore, we sometimes apply the lemma in the context of GCMA's even though it is not phrased in this context.

\subsection{Temporal Logic}

In the following, it is understood that temporal logic refers to propositional linear-time future temporal logic where the natural numbers are used as the domain of time. For background on temporal logic, we refer to  \cite{emerson-handbook-1990} and \cite{gabbay-hodkinson-reynolds-handbook-1994}. As we are dealing with automata and formal languages, we use an approach where the atomic formulas stand for symbols of an alphabet rather than propositional variables, but note that both approaches are interchangeable.

Given an alphabet $A$, the set of \emph{temporal formulas over $A$,} denoted $\TL A$, is typically inductively defined by:
\begin{enumerate}[(i)]
\item for every $a \in A$, the symbol $a$ is an element of $\TL A$,
\item if $\phi \in \TL A$, so is $\neg \phi$,
\item if $\phi, \psi \in \TL A$, so are $\phi \vee \psi$ and $\phi \wedge \psi$,
\item if $\phi \in \TL A$, so is $\Next \phi$ (``next $\phi$''),
\item if $\phi \in \TL A$, so are $\Eventually \phi$ and $\Always \phi$ (``eventually $\phi$'' and ``always $\phi$''),
\item if $\phi, \psi \in \TL A$, so are $\Until \phi \psi$ and $\Release \phi \psi$ (``$\phi$ until $\psi$'' and ``$\phi$ releases $\psi$''). 
\end{enumerate}
Often, the operators $\StrictEvent$ (``strictly eventually'') and $\StrictAlw$ (``strictly always'') are part of the syntax of temporal logic; we view them as abbreviations of $\Next\Eventually$ and $\Next\Always$. For instance, $\StrictEvent (a \wedge \StrictAlw \neg b)$ is viewed as $\Next (\Eventually (a \wedge \Next (\Always \neg b)))$. (Obviously, $\Eventually$ and $\Always$ can be viewed as abbreviations of $(a \vee \neg a) \UntilSym$ and $(a \wedge \neg a) \ReleaseSym$, respectively.)

Formulas of $\TL A$ are interpreted in $\omega$-words over $A$. For every such word $u$, we define what it means for a formula to hold in $u$, denoted $u \models \phi$, where we omit the straightforward rules for Boolean connectives:
 \begin{enumerate}[$\bullet$]
\item $u \models a$ if $u(0) = a$,
\item $u \models \Next \phi$ if $u[1,*) \models \phi$, where, as usual, $u[1,*)$ denotes the word $u(1)u(2) \dots$,
\item $u \models \Eventually \phi$ if there exists $i \geq 0$ such that $u[i,*) \models \phi$, similarly, $u \models \Always \phi$ if $u[i,*) \models \phi$ for all $i \geq 0$,
\item $u \models \Until \phi \psi$ if there exists $j \geq 0$ such that $u[j,*) \models \psi$ and $u[i,*) \models \phi$ for all $i < j$, similarly, $u \models \Release \phi \psi$ if there exists $j \geq 0$ such that $u[j,*) \models \phi$ and $u[i,*) \models \psi$ for all $
i \leq j$ or if $u[i,*) \models \psi$ for all $i \geq 0$.
\end{enumerate}
Clearly, a formula of the form $\neg \Eventually \phi$ is equivalent to $\Always \neg \phi$, and a formula of the form $\neg (\Until \phi \psi)$ is equivalent to $\Release {\neg \phi} {\neg \psi}$, which means $\Eventually$ and $\Always$ as well as $\mathsf{U}$ and $\mathsf{R}$ are dual to each other; $\Next$ is self-dual.

Given a $\TL A$ formula $\phi$, we write $\Lan{\phi}$ for the set of $\omega$-words over $A$ where $\phi$ holds, that is, $\Lan \phi = \{u \in A^\omega \colon u \models \phi\}$. This $\omega$-language is called the \emph{language defined by $\phi$.}

Given $\TL A$ formulas $\phi$ and $\psi$, we say $\phi$ and $\psi$ are \emph{equivalent}, denoted $\phi \equiv \psi$, if $\Lan \phi = \Lan \psi$ holds.


\subsection{Negation Normal Form}

In the later sections of this paper, we always assume that LTL formulas can be assumed to be in negation normal form, which means (ii) from above is not used. The reason that we can do so is that $\neg$ can easily be ``pushed in'', as is explained in the following lemma.

\begin{lem}
  Let $A$ be some alphabet, $a \in A$, and $\phi, \psi \in \TL A$. Then:
  \begin{align*}
    \neg a & \equiv \bigvee_{b \in A \setminus \{ a\}} b \enspace, &
    \neg \Next \phi & \equiv \Next \neg \phi  \enspace,\\
    \neg \Eventually \phi & \equiv \Always \neg \phi \enspace, &
    \neg \Always \phi & \equiv \Eventually \neg \phi \enspace,\\
    \neg (\Until \phi \psi) & \equiv \Release {\neg \phi} {\neg \psi}
    \enspace, & 
    \neg (\Release \phi \psi) & \equiv \Until {\neg \phi} {\neg \psi}
    \enspace.
  \end{align*}
\end{lem}

\proof[Proof hints.]
  The proofs of the individual equivalences are straightforward. Only the proof of the second one is not generic in the sense that it fails for finite words, but for infinite words, which we only consider, no problem occurs.
\qed

From a complexity point of view, it is important to note that when a formula is converted to negation normal form, the size of the formula does not increase much and neither does the number of its subformulas: the increase in the length is at most the number of occurrences of alphabet symbols in the formula and the increase in the number of subformulas is at most the number of alphabet symbols. These increases do not have any influence on the upper bounds we prove in later chapters.


\subsection{Fragments of Temporal Logic}

An \emph{operator set} is a subset of the set of all basic temporal operators,  $\{\Next, \Eventually, \StrictEvent, \UntilSym\}$. If $A$ is an alphabet and $O$ an \emph{operator set}, then $\TLFrag A O$ denotes all LTL formulas that can be built from $A$ using Boolean connectives and the operators from $O$. We say a language $L \subseteq A^\omega$ is \emph{$O$-expressible} if there is a formula $\phi \in \TLFrag A O$ such that $\Lan \phi = L$. The \emph{$O$-fragment} is the set of all LTL-formulas $\phi$ such that $\Lan \phi$ is $O$-expressible. 

Observe that several operator sets determine the same fragment: $\{\StrictEvent\}$ and $\{\Eventually, \StrictEvent\}$; $\{\UntilSym\}$ and $\{\Eventually, \UntilSym\}$; $\{\StrictEvent, \UntilSym\}$ and $\{\Eventually, \StrictEvent, \UntilSym\}$; $\{\Next, \Eventually\}$, $\{\Next, \StrictEvent\}$ and $\{\Next,\Eventually, \StrictEvent\}$; $\{\Next, \UntilSym\}$ and every superset of this.

What we are aiming at are decision procedures for each fragment except for the one determined by $\{\StrictEvent, \UntilSym\}$.


\subsection{Ehrenfeucht--Fraïssé Games for Temporal Logic}

The statements of our results (Section~\ref{sec:results}) do not involve Ehrenfeucht--Fraïssé games (EF games), but we use them extensively in our proofs. We make use of them in Section~\ref{sec:char-strict-fragm}.

In the following, we recall the basics of EF games for temporal logic, see \cite{etessami-wilke-ic-2000} for details.

A play of a temporal logic EF game is played by two players, Spoiler and Duplicator, on two $\omega$-words over some alphabet $A$, say $u$ and $v$. The game is played in rounds, where in every round, Spoiler moves first and Duplicator replies. The basic idea is that Spoiler is trying to reveal a difference between $u$ and $v$ which can be expressed in temporal logic, while Duplicator is trying to show---by somehow imitating the moves of Spoiler---that there is no such difference.

There are different types of rounds, corresponding to the temporal operators considered. We explain the ones that we need:
\begin{enumerate}[$\triangleright$]
\item \emph{$\Next$-round.}  Spoiler chooses either $u$ or $v$, say $v$, and chops off the first letter of $v$, that is, he replaces $v$ by $v[1,*)$. Duplicator does the same for $u$.
\item \emph{$\Eventually$-round.} Spoiler chooses either $u$ or $v$, say $v$, and chops off an arbitrary finite (possibly empty) prefix, that is, he replaces $v$ by $v[i,*)$ for some $i \geq 0$. Duplicator replaces $u$ (the other word) by $u[j,*)$ for some $j \geq 0$.
\item \emph{$\StrictEvent$-round.} Spoiler chooses either $u$ or $v$, say $v$, and chops off an arbitrary non-empty finite prefix, that is, he replaces $v$ by $v[i,*)$ for some $i > 0$. Duplicator replaces $u$ (the other word) by $u[j,*)$ for some $j > 0$.
\end{enumerate}

\noindent Before the first round, $u(0)$ and $v(0)$ are compared. If they are distinct, then this is a win (an early win) for Spoiler. After each round, the same condition is verified, and, again, if the two symbols are distinct, then this is a win for Spoiler. If, by the end of a play, Spoiler hasn't won, then this play is a win for Duplicator. For a fixed $n$, Duplicator wins the $n$-round game, if Duplicator has a strategy to win it.

When only rounds are allowed that correspond to operators in a temporal operator set $O \subseteq \{\Next, \Eventually, \StrictEvent\}$, then we speak of an $O$-game.

The fundamental property of EF games we are going to use is the following, which was essentially proved in \cite{etessami-wilke-ic-2000}.

\begin{thm}
  \label{thm:ew}
  Let $L$ be a language of $\omega$-words over some alphabet $A$ and $O \subseteq \{\Next, \Eventually, \StrictEvent \}$ a temporal operator set.
  Then the following are equivalent:
  \begin{enumerate}[\em(A)]
  \item $L$ is $O$-expressible.
  \item There is some $k$ such that for all words $u, v \in A^\omega$ with $u \in L \NZG v \in L$, Spoiler has a strategy to win the $O$-game on $u$ and $v$ within $k$ rounds.
  \end{enumerate}
\end{thm}


\subsection{From Temporal Logic to Carton--Michel Automata}

Several translations from temporal logic into Büchi and generalized Büchi automata are known, see, for instance, \cite{wolper-vardi-sistla-1983,vardi-wolper-1994,gerth-peled-vardi-wolper-1995}. Here, we follow the ideas of these papers and ``observe'' that the resulting automaton is a GCMA. This is supposed to be folklore,\footnote{Personal communication of the second author with Olivier Carton: the observation can already be found in the notes by Max Michel which he handed over to Olivier Carton in the last millennium.}  but---to the best of our knowledge---has not been made precise yet.

Let $\phi \in \TL A$ and let $\Sub \phi$ denote the set of its subformulas. We define a GCMA $\AutPhi = (A, 2^{\Sub \phi}, I, \cdot, \mathfrak F)$. Our goal is to construct the automaton in such a way that in the unique final run $r$ of this automaton on a given word $u$ the following holds for every $i$ and every $\psi \in \Sub \phi$: 
\begin{align}
  \label{eq:invariant}
  u[i,*) \models \psi \qquad \text{iff} \qquad \psi \in r(i) \enspace.
\end{align}

First, we set $I = \{\Phi \subseteq \Sub \phi \colon \phi \in \Phi\}$, which is motivated directly by (\ref{eq:invariant}).

Second, we define $a \cdot \Phi$ to be the smallest set $\Psi \in 2^{\Sub \phi}$ satisfying the following conditions:
\begin{enumerate}[(i)]
\item if $a \in \Sub \phi$, then $a \in \Psi$,
\item if $\neg b \in \Sub \phi$ and $b \neq a$, then $\neg b \in \Psi$,
\item if $\psi \in \Psi$ and $\chi \in \Psi$, then $\psi \wedge \chi \in \Psi$,
\item if $\psi \in \Psi$ or $\chi \in \Psi$, then $\psi \vee \chi \in \Psi$,
\item if $\psi \in \Phi$, then $\Next \psi \in \Psi$,
\item if $\psi \in \Psi$ or $\Eventually \psi \in \Phi$, then $\Eventually \psi \in \Psi$,
\item if $\psi \in \Psi$ and $\Always \psi \in \Phi$, then $\Always \psi \in \Psi$,
\item if $\chi \in \Psi$ or if $\psi \in \Psi$ and $\Until \psi \chi \in \Phi$, then $\Until \psi \chi \in \Psi$,
\item if $\chi \in \Psi$ and if $\psi \in \Psi$ or $\Release \psi \chi \in \Phi$, then $\Release \psi \chi \in \Psi$.
\end{enumerate}
This definition reflects the ``local semantics'' of temporal logic, for instance, $\Eventually \psi$ is true now if, and only if, $\psi$ is true now or $\Eventually \psi$ is true in the next point in time. Observe, however, that the fulfillment of $\Eventually \psi$ must not be deferred forever, which means that local conditions are not enough to capture the entire semantics of temporal logic. This is taken care of by the final sets.

Third, we list the subsets of $\Sub \phi$ which belong to $\mathfrak F$:
\begin{enumerate}[$\bullet$]
 \item for every formula $\Eventually \psi \in \Sub \phi$, the set $\{\Phi \subseteq \Sub \phi \colon \psi \in \Phi \text{ or } \Eventually \psi \notin \Phi\}$,
 \item for every formula $\Always \psi \in \Sub \phi$, the set $\{\Phi \subseteq \Sub \phi \colon \Always \psi \in \Phi \text{ or } \psi \notin \Phi\}$,\footnote{In the conference version of this paper \cite{preugschat-wilke-2012} we missed this clause.}
 \item for every formula $\Until \psi \chi$, the set $\{\Phi \subseteq \Sub \phi \colon \chi \in \Phi \text{ or } \Until \psi \chi \notin \Phi\}$,
 \item for every formula $\Release \psi \chi$, the set $\{\Phi \subseteq \Sub \phi \colon \Release \psi \chi \in \Phi \text{ or } \chi \notin \Phi\}$.\footnote{See above.}
\end{enumerate}

\begin{exa}[$\phi = \Release a b$]
  After trimming, the automaton $\AutPhi$ for the formula   $\phi = \Release a b$ looks as follows.
  
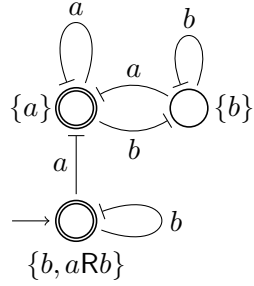
\begin{center}
    \begin{tikzpicture}
      \node[accepting, outer sep=3pt,draw] (0) {};
      \node[left of=0,node distance=6mm] {$\{a\}$};
      \node[draw] (1) [right of=0] {};
      \node[right of=1,node distance=6mm] {$\{b\}$};
      \node[initial, accepting, outer sep=3pt, draw] (2) 
      [node distance=15mm,below of=0] {};
      \node[below of=2,node distance=6mm] {$\{b,\Release a b\}$};
      \path 	  
      (0) edge [bend left] node[lab,above]{$a$} (1)
      (1) edge [bend left] node[lab,below]{$b$} (0);
      \path
      (0) edge node[lab, left]{$a$} (2);
      \path
      (0) edge [loop, looseness=12,out=110,in=70] node[lab,above]{$a$} (0);
      \path
      (1) edge [loop, looseness=11,out=110,in=70] node[lab,above]{$b$} (1);
      \path
      (2) edge [loop, looseness=12,out=20,in=-20] node[lab,right]{$b$} (2);
    \end{tikzpicture}
  \end{center}
  The doubly circled states form the only final state set.
\end{exa}

\begin{prop}
  \label{prop:autphi}
  Let $A$ be an alphabet and $\phi \in \TL A$. Then $\AutPhi$ is a GCMA and $\Lan{\AutPhi} = \Lan{\phi}$.
\end{prop}

\proof We first show that $\AutPhi$ is a GCMA. To this end, let $u$ be an $\omega$-word over~$A$. We show that the word $r$ defined by (\ref{eq:invariant}), for every $i$ and every $\psi \in \Sub \phi$, is a final run on $u$ and the only one. 

  \emph{The $\omega$-word $r$ is a run on $u$.} To see this, let $i \geq 0$ be arbitrary and observe that if we define $\Phi$ and $\Psi$ by $\Phi = \{\psi \in \Sub \phi \colon u[i+1,*) \models \psi\}$ and $\Psi = \{\psi \in \Sub \phi \colon u[i,*) \models \psi\}$, then the implications (i)--(ix) not only hold, but also hold in the opposite direction. That is, $r(i) = u(i) \cdot r(i+1)$ for every $i$, in other words, $r$ is a run on~$u$.

  \emph{The run $r$ is final.} Obvious from the semantics of the temporal operators.
  
  \emph{The run $r$ is the only possible final run.} A proof of this can be carried out along the lines of the proof of Theorem 5.37 in \cite{baier-katoen-2008}, where a variant of the construction from \cite{wolper-vardi-sistla-1983} is presented and proved correct. The only differences between the setting in \cite{baier-katoen-2008} and our setting are the atomic formulas and the set of temporal operators used. In our setting, atomic formulas correspond to letters of an alphabet; in \cite{baier-katoen-2008}, atomic formulas are propositional variables. We work with a larger set of temporal operators. In the proof in \cite{baier-katoen-2008}, it is shown that an anlogue of (\ref{eq:invariant}) holds for any final run, hence the run $r$ is the only one.\qed


\section{General Approach and Individual Results}

This section has two purposes: it explains our general approach and presents the characterizations we have found. 

\subsection{The General Approach}
\label{sec:approach}

To describe our general approach, we first need to explain what we understand by the left congruence of a GCMA. 

Let $\AutA$ be a GCMA. For every $q \in Q$, let $L_q$ denote the set of words $u \in A^*$ such that $uq \in I$. The relation $\LeftCon$ on $Q$, which we call the \emph{left congruence of $\AutA$,} is defined by $q \LeftCon q'$ when $L_q = L_{q'}$. The terminology is justified:

\begin{rem}
  \label{rem:equiv}
  Let $\AutA$ be a GCMA. Then $\LeftCon$ is a left congruence, that is, $uq \LeftCon uq'$ whenever $u \in A^*$ and $q, q' \in Q$ are such that $q \LeftCon q'$.
\end{rem}

In other words, we can define the \emph{left quotient of $\Aut A$} with respect to $\LeftCon$ to be the reverse semi DFA $\Quot {\Aut A} \LeftCon$ given by  
\begin{align}
  \label{eq:1}
  \Quot {\Aut A} \LeftCon = (A, \Quot {Q'} \LeftCon, \Quot I \LeftCon, \circ)  
\end{align}
where 
\begin{enumerate}[$\bullet$]
\item $Q'$ is the set of all states that occur in some final run of $\AutA$ (active states), and
\item $a \circ (\Quot q \LeftCon) = \Quot {(a \cdot q)} \LeftCon$ for all
  $a \in A$ and $q \in Q'$.
\end{enumerate}
As usual, the attribute ``semi'' refers to the fact that this automaton has no final states nor final sets. 

Next, we combine the left congruence of a GCMA with its loops. The \emph{loop language of a state $q$} of a GCMA $\AutA$ is denoted $\LL q$ and defined by
\begin{align}
  \LL q = \bigcup_{q' \LeftCon q} S(q') \enspace,
\end{align}
that is, $\LL q$ contains all loops at $q$ and at congruent states.

Our general approach is to characterize a fragment of LTL as follows. To check whether a given formula $\phi$ is equivalent to a formula in a given fragment, we compute the GCMA $\AutPhi$ and check various conditions on its left quotient and its loop languages. It turns out that this is sufficient; intuitively, the left quotient accounts for the ``finitary fraction'' of $\Lan \AutPhi$, whereas the loop languages account for its ``infinitary fraction''.


\subsection{Characterization of the Individual Fragments}
\label{sec:results}

The formal statement of our main result is as follows.

\begin{thm}
  \label{thm:result}
  Let $A$ be some alphabet, $\phi$ an LTL-formula, and $O$ a temporal operator set as listed in Table~\ref{tab:results}. Then the following are equivalent:
  \begin{enumerate}[\em(A)]
  \item The formula $\phi$ belongs to the $O$-fragment.
  \item The left quotient of $\AutPhi$ and its loop languages satisfy the respective conditions listed in Table~\ref{tab:results}.  (Information on how to read this table follows.)
  \end{enumerate}
\end{thm}

\newcommand{\PatStutter}{%
  \begin{tikzpicture}[baseline=(X.base)]
    \node[circle,draw,fill=gray!50] (A) {};
    \node[circle,draw,fill=gray!50,right of=A] (B) {};
    \node[circle,draw,right of=B] (C) {};
    \draw[|-] (B) -- (C) node[midway,above,lab] (X) {$a$};
    \draw[|-] (A) -- (B) node[midway,above,lab] {$a$};
  \end{tikzpicture}%
}

\newcommand{\PatComm}{%
  \begin{tikzpicture}[baseline=(X.base)]
    \node[circle,draw] (A) {};
    \node[circle,draw,left of=A] (B) {};
    \node[circle,draw,above left of=B] (C1) {};
    \node[circle,draw,fill=gray!50,left of=C1] (C2) {};
    \node[circle,draw,below left of=B] (D1) {};
    \node[circle,draw,fill=gray!50,left of=D1] (D2) {};
    \draw[-|,dashed] (A) -- (B);
    \draw[-|] (B) -- (C1) node[midway,above,lab] {$a$};
    \draw[-|] (C1) -- (C2) node[midway,above,lab] {$b$};
    \draw[-|] (B) -- (D1) node[midway,above,lab] {$b$};
    \draw[-|] (D1) -- (D2) node[midway,above,lab] {$a$};
  \end{tikzpicture}%
}

\newcommand{\PatXF}{%
  \begin{tikzpicture}[baseline=(X.base)]
    \node[circle,draw,fill=gray!50] (A) {};
    \node[circle,draw,fill=gray!50,below of=A] (B) {};
    \node[circle,draw,right of=A] (C) {};
    \node[circle,draw,below of=C] (D) {};
    \draw[-|] (C) -- (A) node[midway,above,lab] (X) {$a$};
    \draw[-|] (D) -- (B) node[midway,above,lab] {$a$};
    \draw[-|,densely dashed] (D) to[bend left] (C);
    \draw[-|,densely dashed] (C) to[bend left] (D);
  \end{tikzpicture}%
}

\newcommand{\PatXandF}{%
  \begin{tikzpicture}[baseline=(X.base)]
    \node[circle,draw,fill=gray!50] (A) {};
    \node[circle,draw,fill=gray!50,right of=A] (B) {};
    \draw[-|,densely dashed] (A) to[loop,in=150,out=210,looseness=12] node[above=1mm] (X) {$x$} (A);
    \draw[-|,densely dashed] (B) to[loop,in=-30,out=30,looseness=12] node[above=1mm] {$x$} (B);
    \draw[-|,densely dashed] (A) to[bend left] (B);
    \draw[-|,densely dashed] (B) to[bend left] (A);
  \end{tikzpicture}%
}

\newcommand{\PatX}{%
  \begin{tikzpicture}[baseline=(X.base)]
    \node[circle,draw,fill=gray!50] (A) {};
    \node[circle,draw,fill=gray!50,right of=A] (B) {};
    \draw[-|,densely dashed] (A) to[loop,in=120,out=60,looseness=12] node[right=2mm] (X) {$x$} (A);
    \draw[-|,densely dashed] (B) to[loop,in=120,out=60,looseness=12] node[right=2mm] {$x$} (B);
  \end{tikzpicture}%
}

\newcommand{\PatXFlabeled}{%
  \begin{tikzpicture}[baseline=-10mm]
    \node[circle,draw,fill=gray!50] (A) {$\bar p$};
    \node[circle,draw,fill=gray!50,below of=A] (B) {$\bar q$};
    \node[circle,draw,right of=A] (C) {$\bar r$};
    \node[circle,draw,below of=C] (D) {$\bar s$};
    \draw[-|] (C) -- (A) node[midway,above,lab] (X) {$a$};
    \draw[-|] (D) -- (B) node[midway,above,lab] {$a$};
    \draw[-|,densely dashed] (D) to[bend left] node[lab,left] {$x$} (C);
    \draw[-|,densely dashed] (C) to[bend left] node[lab,right] {$y$} (D);
  \end{tikzpicture}%
}

\newcommand{\PatStutterlabeled}{%
  \begin{tikzpicture}[baseline=-1mm]
    \node[circle,draw,fill=gray!50] (A) {$\bar p$};
    \node[circle,draw,fill=gray!50,right of=A] (B) {$\bar q$};
    \node[circle,draw,right of=B] (C) {$\bar r$};
    \draw[|-] (B) -- (C) node[midway,above,lab] (X) {$a$};
    \draw[|-] (A) -- (B) node[midway,above,lab] {$a$};
  \end{tikzpicture}%
}

\newcommand{\PatXandFlabeled}{%
  \begin{tikzpicture}[baseline=0mm]
    \node[circle,draw,fill=gray!50] (A) {$\bar p$};
    \node[circle,draw,fill=gray!50,right of=A] (B) {$\bar q$};
    \draw[-|,densely dashed] (A) to[loop,in=150,out=210,looseness=12] node[above=1mm] (X) {$z$} (A);
    \draw[-|,densely dashed] (B) to[loop,in=-30,out=30,looseness=12] node[above=1mm] {$z$} (B);
    \draw[-|,densely dashed] (A) to[bend left] node[above,lab] {$y$} (B);
    \draw[-|,densely dashed] (B) to[bend left] node[below,lab] {$x$} (A);
  \end{tikzpicture}%
}

\newcommand{\PatXlabeled}{%
  \begin{tikzpicture}[baseline=(X.base)]
    \node[circle,draw,fill=gray!50] (A) {$\bar p$};
    \node[circle,draw,fill=gray!50,right of=A] (B) {$\bar q$};
    \draw[-|,densely dashed] (A) to[loop,in=120,out=60,looseness=12] node[right=2mm] (X) {$x$} (A);
    \draw[-|,densely dashed] (B) to[loop,in=120,out=60,looseness=12] node[right=2mm] {$x$} (B);
  \end{tikzpicture}%
}


\begin{table}[t]
  \centering
  \begin{tabular}{ccc}
    \toprule%
    fragment &  left quotient   & loop languages\\\toprule
    $\Next$  & \PatX & no condition\\ 
    \noalign{\smallskip} \midrule
    \multirow{2}{*}{$\Eventually$} & \PatXF & \multirow{2}{*}{1-locally testable}\\
    \noalign{\smallskip} \cline{2-2} \noalign{\smallskip}
    & \PatStutter & \\ 
    \noalign{\smallskip} \midrule\noalign{\smallskip}
    $\StrictEvent$ & \PatXF & 1-locally testable\\ 
    \noalign{\smallskip}\midrule
    $\Next$, $\Eventually$ & \PatXandF & locally testable\\ \midrule
    $\UntilSym$ & \PatStutter & stutter-invariant\\ 
    \noalign{\smallskip} \bottomrule \noalign{\smallskip}
  \end{tabular}
  \caption{Characterizations of the individual fragments of LTL}
  \label{tab:results}
\end{table}

\noindent Conditions on the left quotient of $\AutPhi$ are phrased in terms of ``forbidden patterns'' (also called ``forbidden configurations'' in \cite{cohen-perrin-pin-jcss-1993}). To explain this, let $\AutA = (A, Q, I, \circ)$ be any reverse semi DFA. Its \emph{transition graph}, denoted $\ClosA$, is the $A$-edge-labeled directed graph $(Q, E)$ where $E = \{(a \circ q, a, q) \colon a \in A,  q \in Q\}$. 

Now, the conditions depicted in the second column of Table~\ref{tab:results} are to be read as follows: the displayed graph(s) do not (!) occur as subgraphs of the transition graph of the left quotient of $\AutPhi$, that is, as subgraphs of $\ClosPhi$. Vertices filled gray must be distinct, the others may coincide (even with gray ones); dashed arrows stand for non-trivial paths.

For instance, the condition for the left quotient in the case of the $\{\Next\}$-fragment requires that the following is not true for $\ClosPhi$: there exist distinct states $q$ and $q'$ and a word $x \in A^+$ such that $q = x \circ q$ and $q' = x \circ q'$.

Note that for the $\{\Next\}$-fragment one forbidden pattern consisting of two strongly connected components is listed, whereas for the $\{\Eventually\}$-fragment two forbidden patterns (indicated by the horizontal line) are listed.

The conditions listed in the third column of Table~\ref{tab:results} are conditions borrowed from formal language theory, which we explain in what follows. For a word $u \in A^*$ and $k \geq 0$, we let $\Prfk{u}$, $\Sffxk{u}$, and $\Occk{u}$ denote the set of prefixes, suffixes, and infixes of $u$ of length $\leq k$, respectively. For words $u,v \in A^*$, we write $u \equiv_{k+1} v$ if $\Prfk{u} = \Prfk{v}$, $\OccK  {u} {k+1} = \OccK {v} {k+1}$, and $\Sffxk{u} = \Sffxk{v}$. A language $L$ is called $(k+1)$-locally testable if $u \in L \leftrightarrow v \in L$, whenever $u \equiv_k v$, and it is called \emph{locally testable} if it is $k$-locally testable for some $k$, see~\cite{brzozowski-simon-dm-1973}. 

A language $L \subseteq A^+$ is \emph{stutter-invariant} if $uav \in L \leftrightarrow uaav \in L$ holds for all $a \in A$, $u, v \in A^*$. 


\subsection{Proof techniques}

For each fragment dealt with in Theorem~\ref{thm:result}, we have a separate proof, some of them are similar, others are completely different. In this section, we give a brief overview of our proofs.

For the operator set $\{\Next\}$, the proof is more or less a simple exercise, given that $\{\Next\}$-expressibility means that there is some $k$ such that $u \models \phi$ is determined by $\Prfk u$. 

For the operator sets $\{\Eventually\}$, $\{\StrictEvent\}$, and $\{\Next, \Eventually\}$, we use similar proofs.

For $\{\UntilSym\}$, we use a theorem from~\cite{peled-wilke-ipl-1997}, which says that an LTL formula over some alphabet $A$ is equivalent to a formula in $\TLFrag A {\UntilSym}$ if the language defined by the formula is stutter-invariant, where stutter invariance is defined using an appropriate notion of stutter equivalence on $\omega$-words.

Throughout the next sections, for ease in notation, we often write $\bar q$ for $\Quot q \LeftCon$, where $q$ is a state in $\AutA$. When $u \in A^\omega$, then $u \cdot \infty$ denotes the first state of the unique final run of $\AutA$ on $u$, and $\Inf u = \{a \in A \colon \exists^\infty i (u(i) = a)\}$. For $a \in A$ and $u \in A^*$, $\No a u$ denotes the number of occurrences of $a$ in $u$.


\section{Characterization of the \texorpdfstring{$\{\Next\}$}{{X}}-Fragment}

We start with the characterization of the $\{\Next\}$-fragment, which is  straightforward.

\begin{thm}
\label{thm:tln}
  The following are equivalent for a given trim GCMA $\AutA$:
  \begin{enumerate}[\em(A)]
  \item $L(\Aut A)$ is $\Next$-expressible.
  \item The transition graph $\ClosQA$ does not have a subgraph of the following form (in the above sense):
    \begin{align}
      \PatXlabeled \tag{T1}
    \end{align}
  \end{enumerate}
\end{thm}

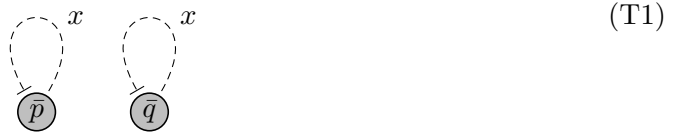
\proof
 (A) implies (B): Let $L(\Aut A)$ be $\Next$-expressible. Let $\phi \in \TLFrag A \Next$ such that $\Lan {\Aut A} = \Lan \phi$. Let $k = \text{length}(\phi)$ where length may be any reasonable function to determine the length of a given formula $\phi$ as a natural number. Obviously for each $w \in \Lan \phi$ and $v \in A^\omega$ the following implication holds: If $\Prfk v = \Prfk w$ then $v \in \Lan \phi$. Let $p \in \bar p$ and $q \in \bar q$. Then there exists $u \in A^*$ with $u \cdot p \in I \NZG u \cdot q \in I$. Let $v,v'\in A^\omega$ such that $p = v \cdot \infty$ and $q = v' \cdot \infty$. Assume that $\ClosQA$ has a subgraph of type (T1). Then $\Prfk {ux^kv} = \Prfk {ux^kv'}$ but $ux^kv \in \Lan {\Aut A} \NZG ux^kv' \in \Lan {\Aut A}$, which is a contradiction.

 We show that (B) implies (A) by contraposition. Assume $L(\Aut A)$ is  not $\Next$-expressible. Then for every natural number $k$ there exist $u,v \in A^\omega$ with $\Prfk u = \Prfk v$ and $u \in \text{L$(\Aut A)$} \NZG v \in L(\Aut A)$. Let $k \geq |Q^2|$ and $u,v$ as described. Let $r$ be the run of $\Aut A$ on $u$ and $s$ be the run of $\Aut A$ on $v$. Note that $r(i) \NotLeftCon s(i)$ for every $i < k$ because $r(0)\in I$ and $s(0) \notin I$. Since $k \geq |Q^2|$ there exist $i<j<k$ with $r(i) = r(j)$ and $s(i) = s(j)$. From $\Prfk u = \Prfk v$ we get $u(i)\dots u(j-1)$ = $v(i)\dots v(j-1)$ and $\ClosQA$ has a subgraph of Type (T1).
\qed


\section{Characterization of the \texorpdfstring{$\{\StrictEvent\}$}{{XF}}-Fragment}
\label{sec:char-strict-fragm}

The second characterization we prove correct is the one of the $\{\StrictEvent\}$-fragment. Since every GCMA can obviously be turned into an equivalent trim GCMA, all GCMA are assumed to be trim subsequently. 

We start with a refined version of Theorem~\ref{thm:result} for the $\{\StrictEvent\}$-fragment.

\begin{thm}
\label{thm:tlSXF}
  The following are equivalent for a given trim GCMA $\AutA$:
  \begin{enumerate}[\em(A)]
  \item $L(\AutA)$ is $\StrictEvent$-expressible.
  \item
    \begin{enumerate}[\em(a)]
    \item The transition graph $\ClosQA$ does not have a subgraph of the following form (in the above sense):
      \begin{align}
        \PatXFlabeled \tag{T2}
      \end{align}
    \item For all $u,v \in A^+$ with $\Occ u =\Occ v$, it holds that $\Loop u \LeftCon \Loop v$.
    \end{enumerate}
    \item
    \begin{enumerate}[\em(a)]
    \item The same as in (B)(a).
    \item
    \begin{enumerate}[\em(i)]
    \item For all $u,v \in A^*,$ $a \in A$, it holds that $\Loop{uav} \LeftCon \Loop{uaav}$.
    \item For all $u,v \in A^*,$ $a,b \in A$, it holds that $\Loop{uabv} \LeftCon \Loop{ubav}$.
    \end{enumerate}
    \end{enumerate}
  \end{enumerate}
\end{thm}

\noindent Observe that (B)(b) means that the loop languages are
1-locally testable. In other words, the above theorem implies that the
characterization of the $\{\StrictEvent\}$-fragment given in
Theorem~\ref{thm:result} is correct.

Before we get to the proof of Theorem~\ref{thm:tlSXF} we provide some more notation and prove some useful lemmas. 

\begin{lem}
  \label{lem:has-patternSXF}
  Assume $\ClosQA$ has a subgraph of type (T2). Then for every $k$ there exist words $u, v \in A^\omega$ such that Duplicator wins the $k$-round $\StrictEvent$-game on $u$ and $v$, but $u \in \Lan{\AutA} \NZG v \in \Lan{\AutA}$.
\end{lem}

\proof
  Assume $\ClosQA$ has a subgraph of type (T2). That is, there are states $\bar p \neq \bar q,\bar r, \bar s$, words $x, y \in A^+$, and a letter $a \in A$ such that $\bar p = a \circ \bar r,\enspace \bar q = a \circ \bar s,\enspace \bar s = y \circ \bar r$ and $\bar r = x \circ \bar s$. We find states $r_0, r_1, \dots$, and $s_0, s_1, \dots$ such that 
  \begin{enumerate}[$\bullet$]
  \item $\bar r_i  = \bar r$ and $\bar s_i = \bar s$ for all $i < \omega$, and
  \item $x \cdot s_i = r_i$ and $y \cdot r_i = s_{i+1}$ for all $i < \omega$.
  \end{enumerate}
  Because $Q$ is a finite set, we find $l > 0$ and $i$ such that $r_i = r_{i+l}$. Since $\AutA$ is trim, we find $v$ such that $v \cdot \infty = r_i$ and $u$ such that $ua \cdot r_i \in I$ iff $ua \cdot s_i \notin I$. This means that $ua(yx)^{lm} v \in L \NZG ua x(yx)^{lm} v \in L$ for all $m \geq 1$. 

  Clearly, if we choose $lm > k$, then the two resulting words cannot be distinguished in the $k$-round $\StrictEvent$-game.
\qed

\begin{lem}
  \label{lem:no-patternSXF}
  Let $\AutA$ be a GCMA such that $\ClosQA$ does not have a subgraph of type (T2). Further, let $r$ and $s$ be the unique final runs of $\Aut A$ on words $u, v \in A^\omega$ and define $\bar r$ and $\bar s$ by $\bar r(i) = \Quot{r(i)}{\LeftCon}$ and $\bar s(i) = \Quot{s(i)}{\LeftCon}$ for all $i < \omega$. 

  If $\bar r(0) \neq \bar s(0)$ and $\Inf {\bar r} \cap \Inf {\bar s} \neq \emptyset$, then Spoiler wins the $k$-round  $\StrictEvent$-game on $u$ and $v$ where $k$ is twice the number of states of $\QuotA$.
\end{lem}

\proof
  In the following, we use SCC as an abbreviation for strongly connected component. In our context, a state which is not reachable by a non-trivial path from itself is considered to be an SCC by itself. For every $i < \omega$, let $R_i$ and $S_i$ be the SCC's of $\bar r(i)$ and $\bar s(i)$ in $\Quot {\Aut A} \equiv_\AutA$, respectively. Observe that because of $\Inf {\bar r} \cap \Inf {\bar s} \neq \emptyset$ there is some $l$ such that the $R_i$'s and $S_j$'s are all the same for $i, j \geq l$.

Let $\mathfrak R = \{R_i \colon i > 0\}$, $\mathfrak S = \{S_i \colon i > 0\}$, $m = |\mathfrak R|-1$, and $n = |\mathfrak S|-1$. We show that Spoiler wins the $\StrictEvent$-game in at most $m+n$ rounds. The proof is by induction on $m+n$. 

\emph{Base case.} Let $m=n=0$. Then $R_1 = S_1$. Because of the absence of (T2), we have $u(0) \neq v(0)$, and Spoiler wins instantly. 

\emph{Induction step.} Note that if $r$ is the unique final run of $\AutA$ on $u$, then $r[i,*)$ is the unique final run of $\AutA$ on $u[i,*)$ for every $i$. 

Let $m+n>0$. If $u(0) \neq v(0)$, then Spoiler wins instantly. If $u(0) = v(0)$, we proceed by a case distinction as follows.

  \emph{Case 1, $R_1 = S_1$.} This is impossible because of the absence of (T2).

  \emph{Case 2, $R_1 \neq S_1$, $R_1 \notin \mathfrak S$.} Since $R_1 \notin \mathfrak S$ and $\Inf {\bar r} \cap \Inf {\bar s} \neq \emptyset$ we have $m>0.$ So there must be some $i \geq 1$ such that $\bar r(i) \in R_1$ and $\bar r(i+1) \notin R_1$. Spoiler chooses the word $u$ and replaces $u$ by $u[i,*)$.

  Now Duplicator has to replace $v$ by $v[j,*)$ for some $j>0$. Since $R_1 \notin \mathfrak S$ we have $\bar r(i) \neq \bar s(j)$ and the induction hypothesis applies.

  \emph{Case 3, $R_1 \neq S_1$, $S_1 \notin \mathfrak R$.} Symmetric to Case 2.

  \emph{Case 4, $R_1 \neq S_1$, $R_1 \in \mathfrak S$, and $S_1 \in \mathfrak R$.} Impossible, because $R_1$ would be reachable from $S_1$ and vice versa, which would mean $R_1$ and $S_1$ coincide.
\qed

\begin{lem}
  \label{lem:eq1}
  Let $\AutA$ be a GCMA. Then the following are equivalent:
  \begin{enumerate}[\em(A)]
    \item For all $u,v \in A^+$ with $\Occ u =\Occ v$, it holds that $\Loop u \equiv_\AutA \Loop v$.
    \item
    \begin{enumerate}[\em(a)]
    \item For all $u,v \in A^*,$ $a \in A$, it holds that $\Loop{uav} \equiv_\AutA \Loop{uaav}$.
    \item For all $u,v \in A^*,$ $a,b \in A$, it holds that $\Loop{uabv} \equiv_\AutA \Loop{ubav}$.
    \end{enumerate}
    \end{enumerate}
\end{lem}

\proof
  That (A) implies (B)  is obvious. For the converse, let $u,v \in A^+$ with $\Occ u = \Occ v$. Let  $\Occ u = \{a_0, a_1, \dots, a_n\}$. Now, we have
 \begin{align*}
   \Loop{u} 
   \LeftCon \Loop{a_0^{\No {a_0} u} a_1^{\No {a_1} u} \dots a_n^{\No {a_n} u}}
   \Loop{\LeftCon a_0^{\No {a_0} v} a_1^{\No {a_1} v} \dots a_n^{\No {a_n} v}}
    \LeftCon \Loop{v} \enspace,
 \end{align*}
 where the first and the last equivalence are obtained by iterated application of~(b), and the second equivalence is obtained by iterated application of~(a).
\qed

In what follows, we need more notation and terminology. A word $u \in A^\omega$ is an \emph{infinite loop} at $q$ if $q = u \cdot \infty$ and $q \in \Inf{r}$ where $r$ is the unique final run of $\AutA$ on $u$.

\paragraph*{Proof of Theorem~\ref{thm:tlSXF}.}
  The implication from (A) to (B)(a) is Lemma~\ref{lem:has-patternSXF}. We prove that (A) implies (B)(b) by contraposition. Assume (B)(b) does not hold, that is, there are $u, v \in A^+$ with $\Occ u = \Occ v$, and $\Loop u \NotLeftCon \Loop v$. Then there exists $x \in A^*$ such that $x \cdot \Loop u \in I \NZG x \cdot \Loop v \in I$, that is, $xu^\omega 	\in L \NZG xv^\omega \in L$. It is easy to see that Duplicator wins the $\StrictEvent$-game on $xu^\omega$ and $xv^\omega$ for any number of rounds, which, in turn, implies $L$ is not $\StrictEvent$-expressible.

  For the implication from (B) to (A), let $n$ be the number of states of $\Quot {\Aut A} \LeftCon$. We show that whenever $u, v \in A^\omega$ such that $u \in L \NZG v \in L$, then Spoiler wins the $2n$-round $\StrictEvent$-game on $u$ and $v$.

  Assume $u, v \in A^\omega$ are such that $u \in L \NZG v \in L$ and let $r$ and $s$ be the unique final runs of $\AutA$ on $u$ and $v$, respectively, and $\bar r$ and $\bar s$ defined as in Lemma \ref{lem:no-patternSXF}. We distinguish two cases.

\noindent\emph{First case, $\Inf u \neq \Inf v$.} Then Spoiler wins within $2$ rounds.
  
\noindent\emph{Second case, $\Inf u = \Inf v$.} Then there are $i,i'$ and $j,j'$ such that
  \begin{enumerate}[$\bullet$]
  \item $\Occ{u[i,j]} = \Occ{v[i',j']}$, 
  \item $u[i,*) \cdot \infty$ is an infinite loop at $\Loop{u[i,j]}$, and
  \item $v[i',*) \cdot \infty$ is an infinite loop at $\Loop{v[i',j']}$.
  \end{enumerate}
  From (B)(b), we conclude $\Loop{u[i,j]} \LeftCon \Loop{v[i',j']}$. As a consequence, $\Inf{\bar r} \cap \Inf{\bar s} \neq \emptyset$. Since $\bar r(0) \neq \bar s(0)$, Lemma~\ref{lem:no-patternSXF} applies: $L$ is $\StrictEvent$-expressible.

  The equivalence between (B) and (C) follows directly from Lemma~\ref{lem:eq1}.\qed


\section{Characterization of the \texorpdfstring{$\{\Eventually\}$}{{F}}-Fragment}

The characterization of the $\{\Eventually\}$-fragment is similar to the one of the $\{\StrictEvent\}$-fragment, but a little more complicated.

\begin{thm}
\label{thm:tle}
  The following are equivalent for a given trim GCMA $\AutA$:
  \begin{enumerate}[\em(A)]
  \item$L(\Aut A)$ is $\Eventually$-expressible.
  \item
    \begin{enumerate}[\em(a)]
    \item The transition graph $\ClosQA$ does not have a subgraph of the following form (in the above sense):
      
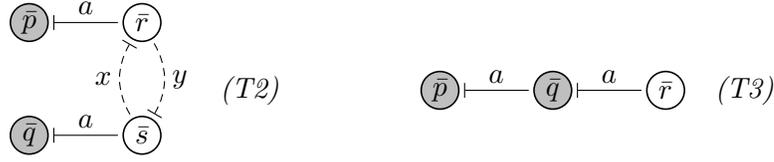
\hfill\PatXFlabeled \quad(T2)\hfill\PatStutterlabeled \quad (T3)\hbox to 50 pt{\hfill}\bigskip

    \item For all $u,v \in A^+$ with $\Occ u =\Occ v$ it holds that $\Loop u \equiv_\AutA \Loop v$.
    \end{enumerate}
    \item
    \begin{enumerate}[\em(a)]
    \item The same as in (B)(a).
    \item
    \begin{enumerate}[\em(i)]
    \item For all $u,v \in A^*,$ $a \in A$ it holds that $\Loop{uav} \equiv_\AutA \Loop{uaav}$.
    \item For all $u,v \in A^*,$ $a,b \in A$ it holds that $\Loop{uabv} \equiv_\AutA \Loop{ubav}$.
    \end{enumerate}
    \end{enumerate}
  \end{enumerate}
\end{thm}

\noindent As seen above (B)(b) means that the loop languages are
1-locally testable. In other words, the above theorem implies that the
characterization of the $\{\Eventually\}$-fragment given in
Theorem~\ref{thm:result} is correct.

Before we turn to the proof we will state some useful lemmas:

\begin{lem}
  \label{lem:has-pattern}
  Assume $\ClosQA$ has a subgraph of type (T2) or (T3). Then for every $k$ there exist words $u, v \in A^\omega$ such that Duplicator wins the $k$-round $\Eventually$-game on $u$ and $v$, but $u \in \Lan{\AutA} \NZG v \in \Lan{\AutA}$.
\end{lem}

\proof
  First, assume $\ClosQA$ has a subgraph of type (T3). That is, there are states $\bar p, \bar q, \bar r \in \Quot Q \equiv_\AutA$ and a symbol $a$ such that $a \circ \bar r = \bar q$, $a \circ \bar q = \bar p$, and $\bar p \neq \bar q$. Let $r \in \bar r$ and define $p$ and $q$ by $q = a \cdot r$ and $p = a \cdot q$. Then $p \in \bar p$ and $q \in \bar q$, because $\equiv_\AutA$ is a left congruence.

 There is some $v \in A^\omega$ such that $v \cdot \infty = r$. Further, since $\bar p \neq \bar q$, there is some $u \in A^*$ such that $u \cdot p \in I \NZG u \cdot q \in I$. In other words, $u a v \in L \NZG u a a v \in L$. Clearly, the two words cannot be distinguished in the $\Eventually$-game. 

  Second, assume $\ClosQA$ has a subgraph of type (T2). That is, there are states $\bar p \neq \bar q,\bar r, \bar s$ words $x, y \in A^+$ and $a \in A$ such that $\bar p = a \circ \bar r,\enspace \bar q = a \circ \bar s,\enspace \bar s = y \circ \bar r$ and $\bar r = x \circ \bar s$. We find states $r_0, r_1, \dots$ and $s_0, s_1, \dots$ such that 
  \begin{enumerate}[(1)]
  \item $\bar r_i  = \bar r$ and $\bar s_i = \bar s$ for all $i$,
  \item $x \cdot s_i = r_i$ and $y \cdot r_i = s_{i+1}$ for all $i$.
  \end{enumerate}
  Because $Q$ is a finite set, we find $l > 0$ and $i$ such that $r_i = r_{i+l}$. In addition, we find $v$ such that $v \cdot \infty = r_i$ and $u$ such that $ua \cdot r_i \in I \NZG ua \cdot s_i \in I$. This means that $ua(yx)^{lm} v \in L \NZG ua x(yx)^{lm} v \in L$ for all $m \geq 1$. 

  Clearly, if we choose $lm \geq k$, then the two resulting words cannot be distinguished in the $k$-round $\Eventually$-game.
\qed

\begin{lem}
  \label{lem:no-pattern}
  Let $\AutA$ be a GCMA such that $\ClosQA$ does not have a subgraph of type (T2) or (T3). Further, let $r$ and $s$ be the unique final runs of $\Aut A$ on words $u, v \in A^\omega$ and define $\bar r$ and $\bar s$ by $\bar r(i) = \Quot{r(i)}{\LeftCon}$ and $\bar s(i) = \Quot{s(i)}{\LeftCon}$ for all $i < \omega$. 

  If $\bar r(0) \neq \bar s(0)$ and $\Inf {\bar r} \cap \Inf {\bar s} \neq \emptyset$, then Spoiler wins the $k$-round  $\Eventually$-game on $u$ and $v$ where $k$ is twice the number of states of $\QuotA$.
\end{lem}

\proof
  Let $R_i$ and $S_j$ be the SCC's of $\bar r(i)$ and $\bar s(j)$ in $\Quot {\Aut A} \equiv_\AutA$, respectively.

  There are $i$ and $j$ such that the SCC's of $\bar r(i')$ and $\bar s(j')$ for $i' \geq i$ and $j' \geq j$ are all the same.
 
  Let $\mathfrak R = \{R_i \colon i > 0\}$, $\mathfrak S = \{S_i \colon i > 0\}$, $m = |\mathfrak R|$, and $n = |\mathfrak S|$. We show that Spoiler wins the game in at most $m+n$ rounds. The proof is by induction on $m+n$. If $u(0) \neq v(0)$ Spoiler wins instantly. Otherwise, we distinguish several cases.

  \emph{Case 1, $R_1 = S_1$.} This is impossible because of the absence of (T2).

  \emph{Case 2, $R_1 \neq S_1$, $R_1 \notin \mathfrak S$.} Since $R_1 \notin \mathfrak S$ and $\Inf {\bar r} \cap \Inf {\bar s} \neq \emptyset$ we have $m>1.$ So there must be some $i \geq 1$ such that $\bar r(i) \in R_1$ and $\bar r(i+1) \notin R_1$. Spoiler chooses the word $u$ and replaces $u$ by $u[i,*)$.

  If $u(i) \neq v(0)$ Duplicator has to replace $v$ by $v[j,*)$ for some $j>0$ if she does not want to lose right away. The induction hypothesis applies since $\bar r(i) \in R_1 \not\in \mathfrak S$ and so $\bar r(i) \neq \bar s(j).$

  If $u(i) = v(0)(= u(0))$, we have to show that $\bar r(i) \neq \bar s(0)$ to be able to apply the induction hypothesis. Assume that $\bar r(i) = \bar s(0)$. Since $\bar r(i) \in R_1 \not\in \mathfrak S$ and $\bar s(1) \in S_1$, we have $\bar s(0) \neq \bar s(1)$ and $\bar s(0) = v(0)\circ \bar s(1)$, i.\,e. $\bar r(i) = v(0)\circ \bar s(1)$ and $\bar r(i) \neq \bar s(1)$. The absence of (T3) leads to $\bar r(i) = v(0)\circ \bar r(i)$ and the absence of (T2) leads to $\bar r(0) = v(0)\circ \bar r(i)$. We get $\bar r(0) = \bar r(i) = \bar s(0)$---a contradiction.

  \emph{Case 3, $R_1 \neq S_1$, $S_1 \notin \mathfrak R$.} Symmetric to Case 2.

  \emph{Case 4, $R_1 \neq S_1$, $R_1 \in \mathfrak S$, and $S_1 \in \mathfrak R$.} Impossible, because $R_1$ would be reachable from $S_1$ and vice versa, which would mean $R_1$ and $S_1$ coincide.
\qed

\paragraph*{Proof of Theorem~\ref{thm:tle}.}
  That (A) implies (B)(a) follows from Lemma~\ref{lem:has-pattern} by contraposition.

  We prove that (A) implies (B)(b) by contraposition. Assume (B)(b) does not hold. Then there are $u, v \in A^+$ with $\Occ u = \Occ v$ and $\Loop u \not\equiv_\AutA \Loop v$. Then there exists $x \in A^*$ such that $x \cdot \Loop u \in I \NZG x \cdot \Loop v \in I$, that is, $xu^\omega \in L \NZG xv^\omega \in L$. Now it is easy to see that Duplicator wins the $\Eventually$-game on $xu^\omega$ and $xv^\omega$ for any number of rounds, which, in turn, implies $L$ is not $\Eventually$-expressible.

  For the implication from (B) to (A), let $n$ be the number of states of $\Quot {\Aut A} \equiv_\AutA$. We show that whenever $u, v \in A^\omega$ such that $u \in L \NZG v \in L$ Spoiler wins the $2n$-round $\Eventually$-game on $u$ and $v$.

  Assume $u, v \in A^\omega$ are such that $u \in L$ and $v \notin L$. We distinguish two cases.

\noindent\emph{First case, $\Inf u \neq \Inf v$.} Then Spoiler wins within at most $2$ rounds.
  
\noindent\emph{Second case, $\Inf u = \Inf v$.} Then there are $i,i'$ and $j,j'$ such that
  \begin{enumerate}[$\bullet$]
  \item $\Occ{u[i,j]} = \Occ{v[i',j']}$,
  \item $u[i,*) \cdot \infty$ is an infinite loop at $\Loop{u[i,j]}$, and
  \item $v[i',*) \cdot \infty$ is an infinite loop at $\Loop{v[i',j']}$.
  \end{enumerate}
  From (B)(b), we conclude $\Loop{u[i,j]} \equiv_\AutA \Loop{v[i',j']}$. As a consequence, Lemma~\ref{lem:no-pattern} applies: $L$ is $\Eventually$-expressible.

  The equivalence between (B) and (C) follows directly from Lemma~\ref{lem:eq1}.\qed


\section{Characterization of the \texorpdfstring{$\{\Next, \Eventually\}$}{{X,F}}-Fragment}

The correctness proof for the characterization of $\{\Next, \Eventually\}$-fragment follows the one for the $\{\Eventually\}$-fragment. We begin with a theorem corresponding to Theorems~\ref{thm:tle} and \ref{thm:tlSXF}.

\begin{thm}
\label{thm:tlne}
  The following are equivalent for a given trim GCMA $\AutA$:
  \begin{enumerate}[\em(A)]
  \item $L(\Aut A)$ is $\Next\Eventually$-expressible.
  \item
    \begin{enumerate}[\em(a)]
    \item The transition graph $\ClosQA$ does not have a subgraph of the following form (in the above sense):
      \begin{align}
        \PatXandFlabeled \tag{T4}
      \end{align}
    \item For some natural $k$ and all $u,v \in A^+$ with $u \equiv_{k+1} v$ we have $\Loop u \equiv_\AutA \Loop v$, i.\,e., for every $q \in Q$, the set $\bigcup_{q' \equiv_\AutA q} S(q')$ is locally testable.
    \end{enumerate}
  \end{enumerate}
\end{thm}

\noindent Before we turn to the proof we will again state some useful lemmas:

\begin{lem}
  \label{lem:has-pattern2}
  Assume the transition graph $\ClosQA$ has a subgraph of type (T4). Then for every $k$ there exist words $u, v \in A^\omega$ such that Duplicator wins the $k$-round $\Next\Eventually$-game on $u$ and $v$, but $u \in L \NZG v \in L$.
\end{lem}

\proof
  Assume (T4) occurs in $\ClosQA$. First observe that for every state $p_i, q_j \in Q$ with $\bar{p_i} = \bar p$ and $\bar{q_i} = \bar q$ and every $l$ it holds that
\[
    z^l \circ \bar p_i = \bar p \enspace, z^l \circ \bar q_j = \bar q \enspace,\bar p =  x \circ \bar q_j \enspace, \bar q =  y \circ \bar p_i \enspace.
\]

Then, observe that for every state $r$ there exist $k$ and $l>0$ such that $z^k \cdot r = z^{k+l} \cdot r$. Since $k$ can be replaced by any larger number and $l$ by any multiple of $l$, we can assume $k$ and $l$ are the same for all states. Let $l$ be fixed with that properties. 
  
Let $p_i \in Q$ with $\bar p_i = \bar p$. Since Q is finite, there exist $j,m$ with $(xz^{2l}yz^{2l})^j\cdot p_i  = (xz^{2l}yz^{2l})^{j+m}\cdot p_i$. It follows easily, that there exist $x',y',z' \in A^+$ and $p',q' \in Q$ with $\enspace p' \neq q', \enspace \bar p' = \bar p,\enspace \bar q' = \bar q$ and $$p' = z' \cdot p',\enspace q' = z' \cdot q',\enspace p' = x' \cdot q',\enspace q' = y' \cdot p',\enspace$$ meaning that $\ClosA$ also has a subgraph of type (T4).

In addition, we find $u \in A^*$ such that $u \cdot p' \in I \NZG u \cdot q' \in I$ and $v \in A^\omega$ such that $p' = v \cdot \infty$ and $q' = y'v \cdot \infty$. This means that $u((z')^nx'(z')^ny')^n(z')^nv \in L \NZG uy'((z')^nx'(z')^ny')^n(z')^nv \in L$ for all $n \geq 1$.

Clearly, if we choose $n > k$, then the two resulting words cannot be distinguished in the $k$-round $\Next\Eventually$-game.
\qed

\begin{lem}
\label{lem:no-pattern2}
  Let $\AutA$ be a GCMA such that $\ClosQA$ does not have a subgraph of type (T4). Further, let $r$ and $s$ be the unique final runs of $\Aut A$ on words $u, v \in A^\omega$ and define $\bar r$ and $\bar s$ by $\bar r(i) = \Quot{r(i)}{\LeftCon}$ and $\bar s(i) = \Quot{s(i)}{\LeftCon}$ for all $i < \omega$.  
 
   Assume  $\bar r(0) \neq \bar s(0)$ and $\Inf {\bar r} \cap \Inf {\bar s} \neq \emptyset$. Let $$\mathfrak Q =\{Q_i \subseteq \Quot Q \equiv_\AutA:Q_i \text{ is an SCC of } \Quot Q \equiv_\AutA\}$$ and $$ K = 2\sum_{Q_i \in \mathfrak Q} {|Q_i|^2} + 2. $$
Then Spoiler wins the $K$-round $\Next \Eventually$-game on $u$ and $v$.
\end{lem}

\proof

Let $R_i$ and $S_j$ be the SCC's of $\bar r(i)$ and $\bar s(j)$ in $\Quot {\Aut A} \equiv_\AutA$, respectively.

There are $i$ and $j$ such that the SCC's of $\bar r(i')$ and $\bar s(j')$ for $i' \geq i$ and $j' \geq j$ are all the same.

Let $\mathfrak R = \{R_i \colon i > 0\}$, $\mathfrak S = \{S_i \colon i > 0\}$, $m = |\mathfrak R|$, and $n = |\mathfrak S|$. We show that Spoiler wins the game in at most $K$ rounds. The proof is by induction on the induction parameter
  \begin{align*}
    \sum_{R \in \mathfrak R} |R|^2 + \sum_{S \in \mathfrak S}
    |S|^2 + [\bar r(0) \notin R_1] + [\bar s(0) \notin S_1] \enspace.
  \end{align*}
  Here, $[r(0) \notin R_1]$ yields $1$ if the condition is true and $0$ otherwise, similarly for $[s(0) \notin S_1]$. Adding $[\bar r(0) \notin R_1]$ and $[\bar s(0) \notin S_1]$ makes sure that if $\bar r(0) \notin R_1$ or $\bar s(0) \notin S_1$, then a $\Next$-move decreases the induction parameter. If $u(0) \neq v(0)$ Spoiler wins instantly. Otherwise, we distinguish several cases.

  \emph{Case 1, $R_1 = S_1$.} Let $c = |R_1|^2$. Spoiler plays $c$ $\Next$-rounds. If Spoiler does not win in these rounds, then $\bar r(c+1) \notin R_1$ or $\bar s(c+1) \notin S_1$ because $\ClosQA$ does not have a subgraph of type (T4)
, and, since $\Aut A$ is co-deterministic, $\bar r(c) \neq \bar s(c)$. The induction hypothesis applies.

  \emph{Case 2, $R_1 \neq S_1$, $R_1 \notin \mathfrak S$.} Then $m > 1$ and there must be some $i \geq 1$ such that $r(i) \in R_1$ and $r(i+1) \notin R_1$. We distinguish two subcases.

  \emph{Subcase 2.a, $\bar r(i) = \bar s(0)$.} Spoiler plays a $\Next$-round, which means Spoiler wins right away or the game proceeds with words such that their runs start in $\bar r(1)$ and $\bar s(1)$, respectively. The induction hypothesis applies, as $s(0) \notin S_1$, see above. 

  \emph{Subcase 2.b, $\bar r(i) \neq s(0)$.} Spoiler plays an $\Eventually$-round, chooses the word $u$, and replaces $u$ by $u[i,*)$. The induction parameter decreases by this, because $|R_1| \geq 2$ or $\bar r(0) \notin R_1$. If Duplicator chooses to not change $v$, then the resulting runs start with $\bar r(i)$ and $\bar s(0)$, which are distinct. If not, then the runs start with $\bar r(i)$ and $\bar s(j)$ for some $j \geq 1$, which are states that do not belong to the same SCC and, hence, are distinct.
 
  \emph{Case 3, $R_1 \neq S_1$, $S_1 \notin \mathfrak R$, and $n > 1$.} Symmetric to Case 2.

  \emph{Case 4, $R_1 \neq S_1$, $R_1 \in \mathfrak S$, and $S_1 \in \mathfrak R$.} Impossible, because $R_1$ would be reachable from $S_1$ and vice versa, which would mean $R_1$ and $S_1$ coincide.
\qed

%
%

For $\omega$-words $u$ and $v$ and a natural number $k$, we write $u \approx_{k+1} v$ if $\Prfk u = \Prfk v $ and $\OccK u {k+1} = \OccK v {k+1} = \InfK u {k+1} = \InfK v {k+1}$. 

\begin{rem}\label{rem:approx2}\hfill
  \begin{enumerate}[(1)]
  \item $\approx_k$ is an equivalence relation.
  \item If $u \approx_{k+1} v$, then $u$ and $v$ cannot be distinguished by the $k$-round $\Eventually\Next$-game.
  \end{enumerate}
\end{rem}

We can finally turn to the correctness proof of our characterization.

\paragraph*{Proof of Theorem~\ref{thm:tlne}.}
 That (A) implies (B)(a) follows from Lemma~\ref{lem:has-pattern2} by contraposition.

  We prove that (A) implies (B)(b) by contraposition. Assume (B)(b) does not hold. Let $k$ be a natural number. There are $u, v \in A^+$ with $u \equiv_{k+1} v$ and $\Loop u \not\equiv_\AutA \Loop v$. Then there exists $x \in A^*$ such that $x \cdot \Loop u \in I \NZG x \cdot \Loop v \in I$, that is, $xu^\omega 	\in L \NZG xv^\omega \in L$. Remark~\ref{rem:approx2} implies Duplicator wins the $k$-round $\Next\Eventually$-game on $u^\omega$ and $v^\omega$ because of $u^\omega \approx_{k+1} v^\omega$. But this implies Duplicator wins the $\Next\Eventually$-game on $xu^\omega$ and $xv^\omega$, which, in turn, implies $L$ is not $\Next\Eventually$-expressible.

 For the implication from (B) to (A), let $K$ be as in Lemma~\ref{lem:no-pattern2}.

We show that whenever $u, v \in A^\omega$ such that $u \in L \NZG v \in L$ Duplicator wins the $\max\{K,2+k\}$-round $\Next\Eventually$-game on $u$ and $v$.

  Assume $u, v \in A^\omega$ are such that $u \in L$ and $v \notin L$. We distinguish two cases.

\noindent\emph{First case, $\Infk u \neq \Infk v$.} Then Spoiler wins within at most $2+k$ rounds.
  
\noindent\emph{Second case, $\Infk u = \Infk v$.} Then there are $i,i'$ and $j,j'$ such that
  \begin{enumerate}[$\bullet$]
  \item $u[i,j] \equiv_k v[i',j']$,
  \item $u[i,*) \cdot \infty$ is an infinite loop at $\Loop{u[i,j]}$, and
  \item $v[i',*) \cdot \infty$ is an infinite loop at $\Loop{v[i',j']}$.
  \end{enumerate}
  From (B)(b), we conclude $\Loop{u[i,j]} \equiv_\AutA \Loop{v[i',j']}$. As a consequence, Lemma~\ref{lem:no-pattern2} applies: $L$ is $\Next\Eventually$-expressible.\qed


\section{Characterization of the \texorpdfstring{$\{ \UntilSym \} $}{{U}}-Fragment}

As mentioned above, the proof for the characterization of the $\{ \UntilSym \} $-Fragment uses a different approach. We begin by stating the result as a theorem.

\begin{thm}
\label{thm:tlu}
  The following are equivalent for a given trim GCMA $\AutA$:
  \begin{enumerate}[\em(A)]
  \item $L(\Aut A)$ is $\UntilSym$-expressible.
  \item \begin{enumerate}[\em(a)]
      \item $L(\Aut A)$ is $\TL A$-definable.
      \item The transition graph $\ClosQA$ does not have a subgraph of the following form (in the above sense):
        \begin{align*}
          \PatStutterlabeled \tag{T3}
        \end{align*}
      \item For all $u,v \in a^*, a \in A: \Loop {uav} \LeftCon \Loop {uaav}$, i.\,e., for every $q \in Q$, the set $\displaystyle\bigcup_{q' \equiv_\AutA q} S(q')$ is stutter-invariant.
    \end{enumerate}
  \end{enumerate}
\end{thm}

\noindent The definition of \emph{stutter-invariance} for languages of $\omega$-words is a little different to the one for finite words. We use the definition from~\cite{peled-wilke-ipl-1997}.
Two $\omega$-words $u$ and $v$ over an alphabet $A$ are called \emph{stutter-equivalent} iff there are two infinite sequences $0 = i_0 < i_1 < i_2 < \dots$ and $0 = j_0 < j_1 < j_2 < \dots$ such that for every $k \geq 0$ $u(i_k)=u(i_k+1)= \dots =u(i_{k+1}-1) = v(j_k)=v(j_k+1)= \dots =v(j_{k+1}-1)$. With the notion of stutter-equivalence we define stutter-invariance for $\omega$-languages. 
An $\omega$-Language $L$ over an alphabet $A$ is said to be stutter-invariant iff for each pair $u,\,v$ of stutter-equivalent words we have $u \in L \leftrightarrow v \in L$

For the proof of the above theorem, we need a theorem from~\cite{peled-wilke-ipl-1997} which reads as follows.

\begin{thm}
\label{thm:tlus}
 A $\TL A$-definable $\omega$-language $L \subseteq A^\omega$ is $\UntilSym$-expressible if and only if $L$ is stutter-invariant.
\end{thm}

\paragraph*{Proof of Theorem~\ref{thm:tlu}.}
 (A) implies (B)(b): Let $\Lan {\Aut A}$ be $\UntilSym$-expressible. By Theorem~\ref{thm:tlus} $\Lan {\Aut A}$ is stutter-invariant. Assume $\ClosQA$ has a subgraph of type (T3). Then there exist $u \in A^*, p \in \bar p$ and $q \in \bar q$ with $u \cdot p \in I \NZG u \cdot p \in I$. Since $\Aut A$ is trim, there exists $v \in A^\omega$ with $v \cdot \infty = q$. So we have $uav \in \Lan {\Aut A} \NZG uaav \in \Lan {\Aut A}$ which means $\Lan {\Aut A}$ is not stutter-invariant --- a contradiction.

 (A) implies (B)(c) by contraposition: Assume there are $u,v \in A^*, a \in A$   with $\Loop {uav} \NotLeftCon \Loop {uaav}$. then there exists $w \in A^*$ with $w \cdot \Loop {uav} \in I \NZG w \cdot \Loop {uaav} \in I$. Hence $w(uav)^\omega \in \Lan {\Aut A} \NZG w(uaav)^\omega \in \Lan {\Aut A}$ and so $\Lan {\Aut A}$ is not stutter-invariant.

 To prove the implication from (B) to (A) we have to show that $\Lan {\Aut A}$ is $\TL A$-definable and stutter-invariant. Then we can apply Theorem~\ref{thm:tlus} and the proof is complete.

 First we show, that $\Lan {\Aut A}$ is stutter-invariant. Let $w \in \Lan {\Aut A}$ with the unique final run $r$ and $\bar r$ the factorization of $r$ as seen above. Let $i \in \Naturals$ with $r(i) \in \Inf r$ and $u = w[0,i)$ and $v = w[i,*)$. Then $r(i) = v \cdot \infty$. Since the loop languages are stutter-invariant and $r(i) \in \Inf r$, for every $v' \in A^*, a \in A$ and $v'' \in A^\omega$ with $v'av'' = v$ there exists $r'(i) \in Q$ with $r'(i) \LeftCon r(i)$ and $r'(i) = v'aav'' \cdot \infty$ which means $uv'aav'' \in \Lan {\Aut A}$. Since $r(i) \in \Inf r$ this argument can be applied infinitely often at once. The absence of (T3) means that for every $x \in A^*$ and every $a \in A$ the equivalence $ax \cdot r(i) \LeftCon aax \cdot r(i)$ holds. If $u \neq \varepsilon$ let $u' \in A^*, a \in A$ and $u'' \in A^*$ with $u'au'' = u$. Then $u'aau'' \cdot r(i) \in I$ and $u'aau''v \in \Lan {\Aut A}$. So $\Lan {\Aut A}$ is stutter-invariant. \qed


\section{Effectiveness and Computational Complexity}

To conclude, we explain how Theorem~\ref{thm:result} can be used effectively. In general, we have:

\begin{thm}
  \label{thm:effective}
  Each of the fragments listed in Table~\ref{tab:results} is decidable.
\end{thm}

Observe that for the fragment with operator set $\{\Eventually,\UntilSym\}$, this is a result from~\cite{peled-wilke-wolper-tcs-1998}, and for the fragment with operator set $\{\Next,\Eventually\}$, this is a result from~\cite{wilke-habil-1998}.

\paragraph*{Proof of Theorem~\ref{thm:effective}.}
  First, observe that $\AutPhi$ can be constructed effectively. Also, it is easy to derive the left quotient of $\AutPhi$ from $\AutPhi$ itself and DFA's for the loop languages, even minimum-state DFA's for them, simply by using any of the available minimization procedures, for instance, the one described in~\cite{hopcroft}.

  Second, observe that the presence of the listed forbidden patterns can be checked effectively. The reason is as follows. The test for the existence of a path between two states can be restricted to paths of length at most the number of states. The test for the existence of two loops with the same label but distinct starting states (see forbidden patterns for $\{\Next\}$ and $\{\Next,\Eventually\}$) in some semi automaton $\AutA = (A, Q, \delta)$ amounts to searching for a loop in the semi automaton $(A, Q \times Q \setminus \{(q,q) \mid q \in Q\}, \delta')$ with transition function defined by $\delta'((q,q'), a) = (\delta(q, a), \delta(q', a))$. In other words, this amounts to a search in the original automaton restricted to paths of length at most the number of states squared.

  Third, the conditions on the loop languages can be checked effectively. For 1-local testability, this is because a language $L \subseteq A^*$ is not 1-locally testable if, and only if, one of the following conditions holds:
  \begin{enumerate}
  \item There are words $u, v \in A^*$ and there is a letter $a \in A$ such that $uav \in L \NZG uaav \in L$.
  \item There are words $u \in A^*, v \in A^*$ and letters $a,b \in A$ such that $uabv \in L \NZG ubav \in L$.
  \end{enumerate}
  Again, $u$ and $v$ can be bounded in length by the number of states. For local testability, we refer to \cite{kim-mcnaughton-mccloskey-toc-1991}, where it was shown this can be decided in polynomial time. For stutter invariance, remember that a language $L \subseteq A^*$ is not stutter-invariant if, and only if, the first from the above conditions holds. So this can be checked effectively, too. (One could also use the forbidden pattern listed.)\qed
As to the computational complexity of the problems considered, we first note:
\begin{prop}
  \label{prop:lower}
  Each of the fragments listed in Table~\ref{tab:results} is PSPACE-hard.
\end{prop}

\proof
  The proof is an adaptation of a proof for a slightly weaker result given in~\cite{peled-wilke-wolper-tcs-1998}. 

  First, recall that LTL satisfiability is PSPACE-hard for some fixed alphabet~\cite{sistla-clarke-jacm-1985}, hence LTL unsatisfiability for this alphabet is PSPACE-hard, too. Let $A$ denote such an alphabet in the following. 

  Second, let $c$, $d$, and $e$ be three distinct symbols not in $A$, let $C = \{c, d, e\}$, and let $F = A \cup C$. For every $\TL A$-formula $\phi$, set 
  \begin{align*}
    \alpha_\phi = c \wedge \Next c \wedge
    \Next (\Until C {\Always A}) 
    \wedge \Eventually (d \wedge \Next (\Until c d)) 
    \wedge \Until C {(A \wedge \phi)} \enspace,
  \end{align*}
  where $A$ stands for $\bigvee_{a \in A} a$ and $C$ for $c \vee d \vee e$.

  The formula $\alpha_\phi$ is chosen in such a way that for every $u \in F^\omega$ the following are equivalent:
  \begin{enumerate}[$\bullet$]
  \item $u \models \alpha_\phi$,
  \item $u$ can be written as $vw$ with $v \in C⁺$ and $w \in A^\omega$ and such that $v \models c \wedge \Next c$, $v \models \Eventually (d \wedge \Next (\Until c d))$, and $w \models \phi$.
  \end{enumerate}
  From~\cite{peled-wilke-ipl-1997} and \cite{etessami-wilke-ic-2000}, it follows that the set of finite words satisfying $c \wedge \Next c$ and $\Eventually (d \wedge \Next (\Until c d))$ is not expressible in any of the fragments considered. So if $\phi$ is satisfiable, then $\alpha_\phi$ is not expressible in any of the fragments. But if $\phi$ is not satisfiable, then so is $\alpha_\phi$, which means $\alpha_\phi$ is expressible in any of the fragments considered. In other words, $\phi \mapsto \alpha_\phi$ is an appropriate reduction to prove the claim of the proposition.
\qed

Our upper bounds are as follows:

\begin{thm}
  \label{thm:lower}
  The $\{\Next, \Eventually\}$-fragment is in E (exponential time), the other fragments listed in Table~\ref{tab:results} are in PSPACE.
\end{thm}

Observe that the result for the $\{\UntilSym\}$-fragment is not new, but was already obtained in~\cite{peled-wilke-wolper-tcs-1998}.

\proof
  The proof is a refinement of the proof of Theorem~\ref{thm:effective}.

  Observe that each property expressed as forbidden pattern (as used in our characterizations) can not only be checked in polynomial time (which is folklore), it can also be checked non-deterministically in logarithmic space, simply by guessing the paths in questions, even if we are given a GCMA and need to check it on its left quotient. So if we interweave the construction of~$\AutPhi$, which has an exponential number of states, with the non-deterministic logarith\-mic-space tests for the existence of forbidden patterns, we obtain a polynomial-space procedure for testing the conditions on $\ClosPhi$. (This is a standard argument in computational complexity.)

  The situation is more complicated for the conditions on the loop languages. First observe that from the automaton~$\AutPhi$ we can get reverse DFA's of size polynomial in the size of $\AutPhi$ such that every loop language is the union of the languages recognized by these reverse DFA's, which allows us to analyze the loop languages effectively.

We first deal with 1-local testability and stutter invariance and start with the observation that 1.\ and 2.\ from the proof of Theorem~\ref{thm:effective} can be adapted as follows. There are two states $p$ and $q$ in $\AutPhi$ that are not equivalent with respect to $\LeftCon$ and such that one of the following conditions is true:
  \begin{enumerate}[(1)]
  \item There are words $u, v \in A^*$ and there is a letter $a \in A$ such that $uav \in \LL p$ and $uaav \in \LL q$.
  \item There are words $u,v \in A^*$ and letters $a,b \in A$ such that $uabv \in \LL p$ and $ubav \in \LL q$.
  \end{enumerate}
  From this, it follows that we can bound the length of $u$ and $v$ polynomially in the size of $\AutPhi$, which again yields polynomial-space procedures for both, 1-local testability and stutter invariance.

  For (general) local testability, we apply the polynomial-time decision procedure for local testability developed in  \cite{kim-mcnaughton-mccloskey-toc-1991} to the product of the reverse DFA's mentioned above, which yields an exponential-time algorithm altogether.
\qed

We conclude this section with a more general version of Theorem~\ref{thm:effective}:
\begin{cor}
  \label{cor:effective}
  For each of the fragments listed in Table~\ref{tab:results}, the following is decidable. Given an $\omega$-regular language $L$, is $L$ definable in the fragment?
\end{cor}
\proof 
  Given $L$ we can construct effectively a Büchi automaton $\AutA$ which recognizes $L$, see \cite{thomas-1990} for example. In \cite{diekert-gastin-2008} the decidability of the LTL-definability of $\Lan{\AutA}=L$ is shown. Theorem~\ref{thm:cm} yields a GCMA $\Aut{B}$ with $\Lan{\Aut{B}}=\Lan{\AutA}$. Theorem~\ref{thm:effective} completes the proof.
\qed

\section{Open problems}

We would like to state some questions:
\begin{enumerate}[(1)]
\item Our lower and upper bounds for the complexity of the $\{\Next,\Eventually\}$-fragment don't match. What is the exact complexity of this fragment?
\item Clearly, from our proofs it can be deduced that if a formula $\phi$ is equivalent to a formula in a fragment, an equivalent formula can be constructed effectively. What is the complexity of this construction task?
\item It is not difficult to come up with examples where every equivalent formula has exponential size (even exponential circuit size). What is the worst-case blow-up?--- Observe that, in terms of circuit size, there is a polynomial upper bound for the $\{U\}$-fragment, see \cite{etessami-ipl-2000}.
\end{enumerate}


\end{document}